\documentclass[aps,prd,twocolumn,amsmath,showpacs,superscriptaddress,nofootinbib]{revtex4-1}
\input psfig.sty
\def \be {\begin{equation}}

\def \ee {\end{equation}}
\def \bea {\begin{eqnarray}}
\def \eea {\end{eqnarray}}

\usepackage{graphicx}% Include figure files
\usepackage{dcolumn}% Align table columns on decimal point
\usepackage{bm}% bold math
\usepackage{epsfig}

\begin{document}
\title{Constraints on the cosmic distance duality relation with simulated data of gravitational waves from the Einstein Telescope}
\author{Tao Yang} \email{yangtao2017@bnu.edu.cn}
\affiliation{Department of Astronomy, Beijing Normal University, Beijing, 100875, China}

\author{R. F. L. Holanda} \email{holandarfl@fisica.ufrn.br}
\affiliation{Departamento de Fisica Teorica e Experimemntal, Universidade Federal do Rio Grande do Norte, 59078-970, Natal - RN, Brasil}

%\author{J. S. Alcaniz} \email{alcaniz@on.br}
%\affiliation{Observat\'orio Nacional, 20921-400, Rio de Janeiro - RJ, Brasil}
%\affiliation{Physics Department, McGill University, Montreal, QC, H3A 2T8, Canada}

\author{Bin Hu} \email{bhu@bnu.edu.cn}
\affiliation{Department of Astronomy, Beijing Normal University, Beijing, 100875, China}

\date{\today}

\begin{abstract}

The cosmic distance duality relation (CDDR) has been test through several astronomical observations in the last years. This relation establishes a simple equation relating the angular diameter ($D_A$) and luminosity ($D_L$) distances at a redshift $z$, $D_LD_A^{-1}(1+z)^{-2}=\eta=1$. However,  only very recently  this relation has been observationally tested at high redshifts ($z \approx 3.6$) by using luminosity distances from type Ia supernovae (SNe Ia) and gamma ray bursts (GRBs) plus angular diameter distances from strong gravitational lensing (SGL) observations. The results show that no significant deviation from the CDDR validity has been verified. In this work, we test  the potentialities of future luminosity distances from  gravitational waves  (GWs) sources to impose limit on possible departures of CDDR jointly with current SGL observations. The basic advantage of $D_L$ from GWs is being insensitive  to non-conservation of the number of photons. By simulating 600, 900 and 1200  data of GWs using the Einstein Telescope (ET) as reference, we derive limits on $\eta(z)$ function and obtain that the results will be at least competitive with current limits from the SNe Ia $+$ GRBs $+$ SGLs analyses.

\end{abstract}
%\pacs{98.80.-k, 95.36.+x, 98.80.Es}
\maketitle

\section{Introduction}

In recent years, the unprecedented quality and quantity of  astronomical data has thrown cosmology into the age of precision. The main cosmological parameters have been obtained with errors of a few percent through  measurements  of the cosmic background radiation (CMB) anisotropies~\cite{Hinshaw:2012aka,Ade2015}, observations of type Ia supernovae (SNe Ia)~\cite{Suzuki2012,Betoule2014} and measurements of the clustering of galaxies at different stages of the universe evolution~\cite{Cole05,Eisenstein:2005,Percival10,Carvalho:2017tuu} (we refer the reader to \cite{Weinberg2013} for a recent review).

This set of observations has established the $\Lambda$CDM model as the standard scenario. However, there is still no satisfactory understanding or description of the main components of this model, namely, the cold dark matter (CDM) and dark energy ($\Lambda$) fields. This motivates the need to probe the foundations and fundamental hypotheses of the standard cosmology. In recent works, the assumption of homogeneity and isotropy of the Universe on large scales has been probed~\cite{Maartens:2011yx,Clarkson2012, Bengaly:2015xkw,Bengaly:2016wto,Goncalves:2017dzs}, as well as the constancy of the fine structure constant \cite{Webb1999, Webb2013,Holanda2016a,Holanda2016b}  and the CMB temperature evolution standard  law \cite{Luzzi2009,Luzzi2015,Hurier2014}.   The validity of the cosmic distance duality relation (CDDR), which relates the luminosity distance $D_L$ of an object at a given redshift $z$ to its angular diameter distance $D_A$ via $D_LD_A^{-1}(1+z)^{-2}=1$~\cite{Etherington1933,Ellis2007} has also been investigated. The CDDR is the subject of the present work.

The cosmic distance duality relation (CDDR) is the astronomical version of the reciprocity theorem derived by \cite{Etherington1933}, which is based on fact that many geometric properties are invariant when the roles of the source and observer in astronomical observations are transposed. The reciprocity relation holds when photons follow null geodesic and  the geodesic deviation equation is valid, while the CDDR  holds if the reciprocity relation is valid and the number of photon is conserved. Thus, the possibilities of the CDDR violation are: non-conservation of the number of photons or evidence for a non-metric theory of gravity, in which photons do not follow null geodesic \cite{Uzan2004} (see also \cite{Santana:2017zvy}). In the former case, non-conservation of the total number of photons can be related to presence of some opacity source and non-standard mechanisms such as the axion-photon conversion induced by intergalactic magnetic fields \cite{basset2004,avg2009,avg2010,jac2010} or scalar fields with a nonminimal multiplicative  coupling  to the electromagnetic Lagrangian \cite{Hees2014,HolandaBarros2016,HolandaSaulo2017,Holandasimoni2017}. We should note that the violation of CDDR may also be related to the modified gravity theory~\cite{Belgacem:2017ihm}, in which the luminosity distance for GWs can differ from that for electromagnetic signals. In this paper, the main motivation is using the GW instead of the SNe Ia to avoid the no-conservation of photos. However, as one may see,  any observational deviation from the CDDR validity would be an indication for new physics.

Several tests have been proposed in the past years assuming a deformed CDDR relation, such as: $D_LD_A^{-1}(1+z)^{-2}=\eta(z)$. The ideal way to observationally test the CDDR is via independent measurements of intrinsic luminosities and sizes of the same object, without using a specific cosmological model~\cite{HGA2012}. We may quote approaches involving: measurements of the angular diameter distance (ADD) of galaxy clusters, observations of SNe Ia,  estimates of the cosmic expansion $H(z)$ from cosmic chronometers, measurements of the gas mass fraction in galaxy clusters and observations of strong gravitational lensing (SGL)~\cite{5,6,7,8,9,10,11,12,13,13a}. All these tests were performed using different sources for $D_A$ and $D_L$. On the other hand, the Ref.~\cite{HGA2012} proposed a test which uses exclusively $D_L$ and $D_A$ measurements of the very same object (massive galaxy clusters), obtained from their Sunyaev-Zel'dovich and X-ray observations. Also, in order to test the CDDR at high-$z$,  Refs.~\cite{15,16} considered a sample of strong gravitational lensing system  along with the latest gamma-ray burst distance modulus data \cite{grb}. No significant deviation from the CDDR validity was verified, although the results did not completely rule out $\eta \neq 1$. It is worth mentioning that a common limitation of all aforementioned tests  is that, if $\eta \neq 1$ is obtained, the fundamental reason for the departure may not be known since the results of these tests are sensitive to both conditions for the CDDR violation.

On the other hand, the recent detection of an electromagnetic counterpart (GRB 170817A) to the gravitational wave signal (GW170817) from the merger of two neutron stars~\cite{TheLIGOScientific:2017qsa, Monitor:2017mdv, Diaz:2017uch, Cowperthwaite:2017dyu} have opened the possibility to obtain not only important astrophysical information on the emitting sources, but also to probe cosmology~\cite{Cai:2017cbj,DiValentino:2017clw}. This makes the first time that a cosmic event has been viewed in both gravitational waves and light. Thus opens the windows of a long-awaited multi-messenger astronomy. The application in cosmology was first proposed by Schutz in 1986, who showed that it is possible to determine the Hubble
constant from GW observations, by using the fact that GWs from stellar binary systems
encode distance information~\cite{Schutz:1986gp}. Thus
the inspiraling and merging compact binaries consisting of neutron stars (NSs) and black holes (BHs), can be considered analogously as the supernovae (SN) standard candles, namely the {\it standard sirens}. Unlike SNe Ia, from GWs one can measure the luminosity standard siren directly without the need of cosmic distance ladder: standard sirens are self-calibrating. This property can help us  dodge the influence of the conservation of the number of photon on the test of CDDR, which should be considered when one uses the SNe Ia observations. Combining the measurements of the sources' redshitfs from, for example, the electromagnetic (EM) counterpart one can obtain the luminosity distance and redshift relationship.

In this paper, we explore the ability of the gravitational wave detections to constrain a possible departure from the CDDR. The analyses are performed by using luminosity distances of simulated data of gravitational waves while the  angular diameter distances are from 118 strong gravitational lensing systems with redshift ranges $0.075 \leq z_l \leq 1.004$ and $0.20 \leq z_s \leq 3.60$. We estimate the constraints on $\eta(z)$ parameterisations from the simulation of 600, 900 and 1200  data of GWs using the Einstein Telescope (ET) as reference. The ET is a third-generation ground-based detector of GWs being ten times more sensitive in amplitude than the advanced ground-based detectors, covering the frequency range of $1-10^4$ Hz. We use Gaussian process (GP) to obtain the luminosity distances for each SGL system  from simulated GW data. In order to compare our results with the previous one of Refs.~\cite{15,16} , which  tested the CDDR using SGL, SNe Ia and GRBs, we consider threes  functional forms of $\eta(z)$: $\eta(z)=1+\eta_0 z$, $\eta(z)=1+\eta_0 z/(1+z)$ and $\eta(z)=1+\eta_0\ln(1+z)$  (if $\eta_0=0$ the validity is verified). As  result, we obtain that future results from GWs will be at least  competitive with current limits from current analyses.

%\begin{figure*}
%\centering
%\includegraphics[width=0.47\textwidth]{fig1a.eps}
%\hspace{0.3cm}
%\includegraphics[width=0.47\textwidth]{fig1b.eps}
%\caption{In Fig. (1a) we plot the SNe Ia from Union2.1 compilation (black stars) and the points used in our analyses (red circles, see Eq.\ref{eq:dlsigdl}). In Fig. (1b) we plot the 82 gas mass fractions calculated from Ref.\cite{has}. We ruled out 9 galaxy clusters  from original 91 data points due to they do not have SNe Ia pairs with $\Delta z\leq 0.005$.}
%\end{figure*}
\section{Testing the CDDR with strong gravitational lensing}

Recently, some works used strong gravitational lensing observations to test the CDDR in a cosmological model independent approach \cite{13a,15,16}. For a flat universe, the approach is as follows.

Let us consider  the following observational quantity obtained from strong gravitational lensing systems:
\begin{equation}
\label{D}
D={\frac{D_{A_{ls}}} {D_{A_s}}}=\frac{{\theta}_E c^2}{4{\pi} \sigma^2_{SIS}},
\end{equation}
where the subindices (ls), (s) and (l) correspond to lens-source, source and lens, respectively. The comoving distance $r_{ls}$ can be written  as%~\cite{Bal}
\begin{equation}
r_{ls}=r_s-r_l.
\end{equation}
Using $r_s=(1+z_s)D_{A_s}$, $r_l=(1+z_l)D_{A_l}$  and $r_{ls}=(1+z_s)D_{A_{ls}}$, one may find
\begin{equation}
\label{d2}
D= 1 - \frac{(1+z_l)D_{A_{l}}}{(1+z_s)D_{A_{s}}}.
\end{equation}
Finally, if one considers the deformed CDDR, such as, $D_L D_A^{-1}(1+z)^{-2} = \eta(z) $, the above expression takes the form
\begin{equation}
\label{d3}
\frac{(1+z_s)\eta(z_s)}{(1+z_l)\eta(z_l)}= (1-D)\frac{D_{L_s}}{D_{L_l}}.
\end{equation}
%%%%%%%%%%%%%%%%%%%%

\begin{figure*}
\centering
\includegraphics[width=0.3\textwidth]{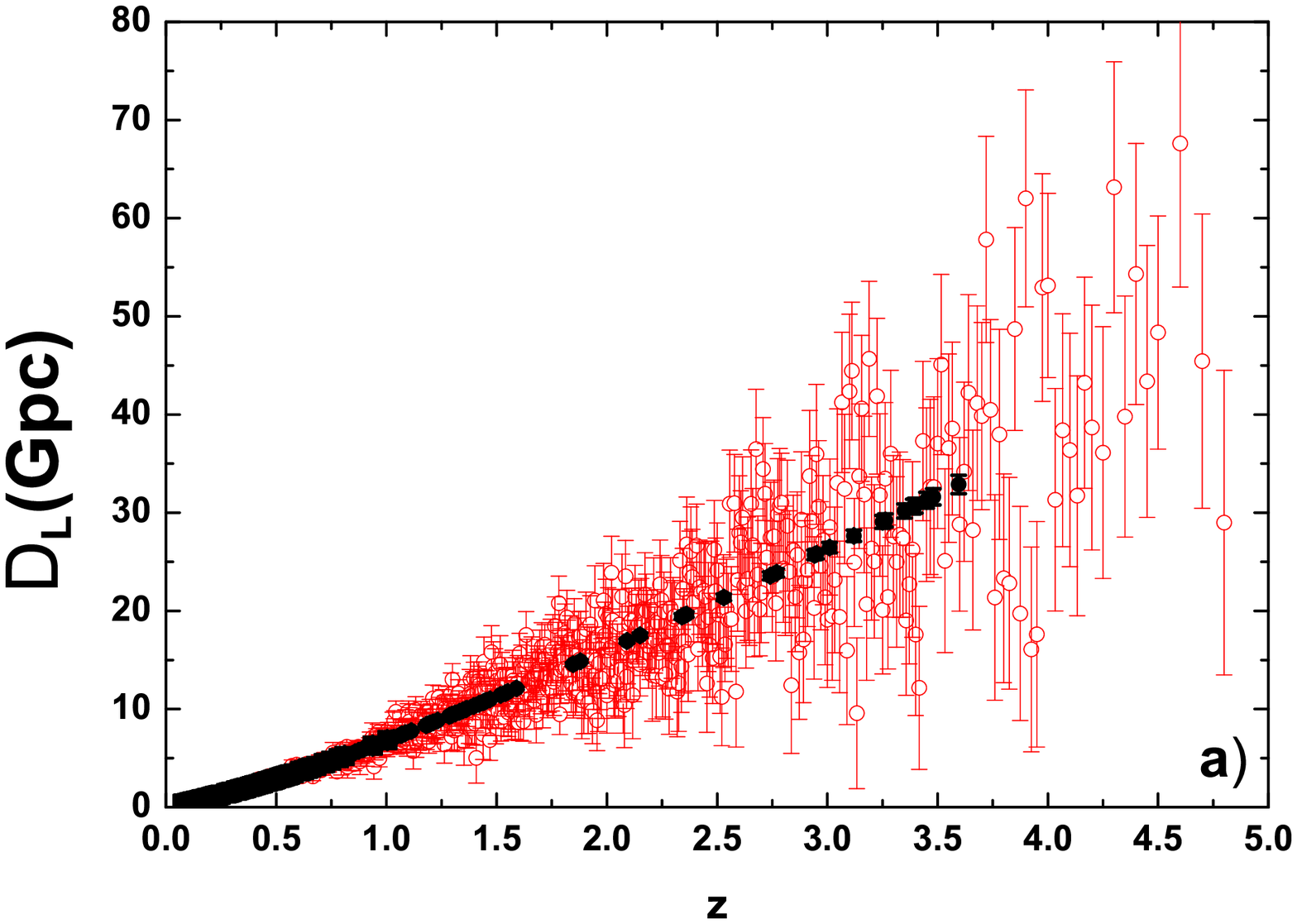}
\hspace{0.3cm}
\includegraphics[width=0.3\textwidth]{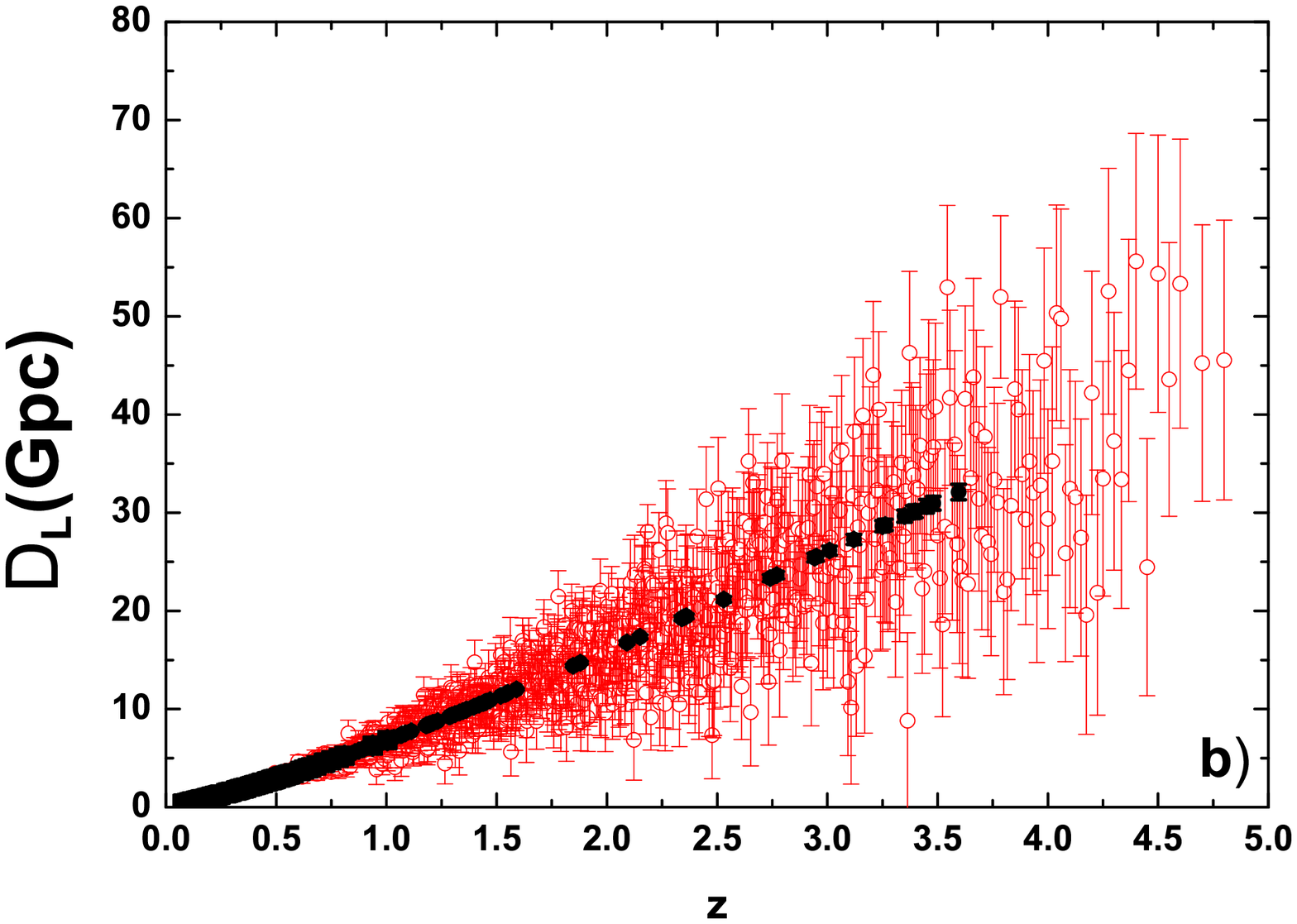}
\hspace{0.3cm}
\includegraphics[width=0.3\textwidth]{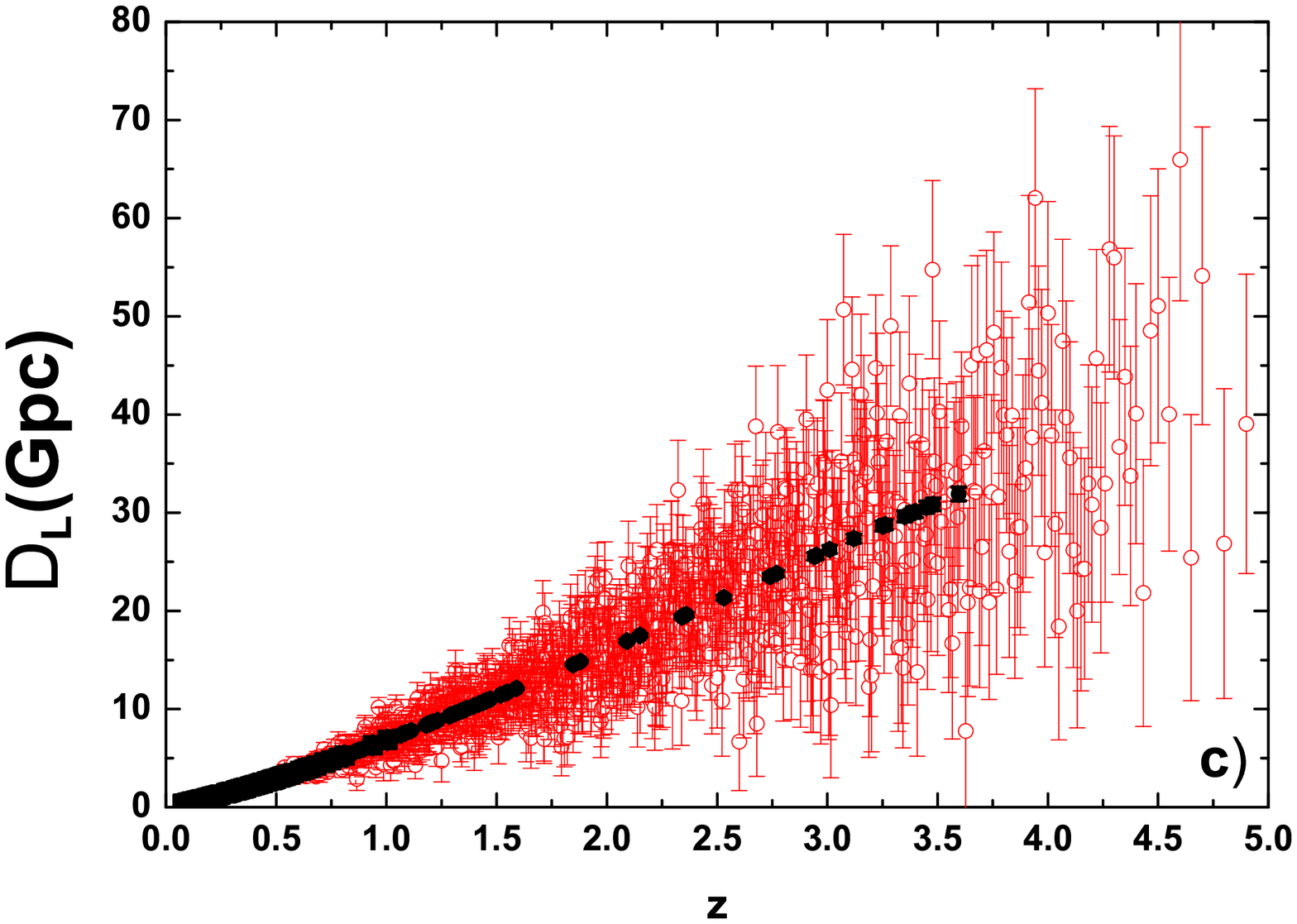}
\caption{ In all figures the luminosity distances and errors obtained from the simulated GW data for each SGL system are shown as red open circles. The luminosity distances and errors obtained for each SGL system by using Gaussian Process are shown as black filled circles. From left to right the number of the simulated GW data is 600, 900, and 1200.}
\label{fig:plotdl}
\end{figure*}

\begin{figure*}
\centering
\includegraphics[width=0.3\textwidth]{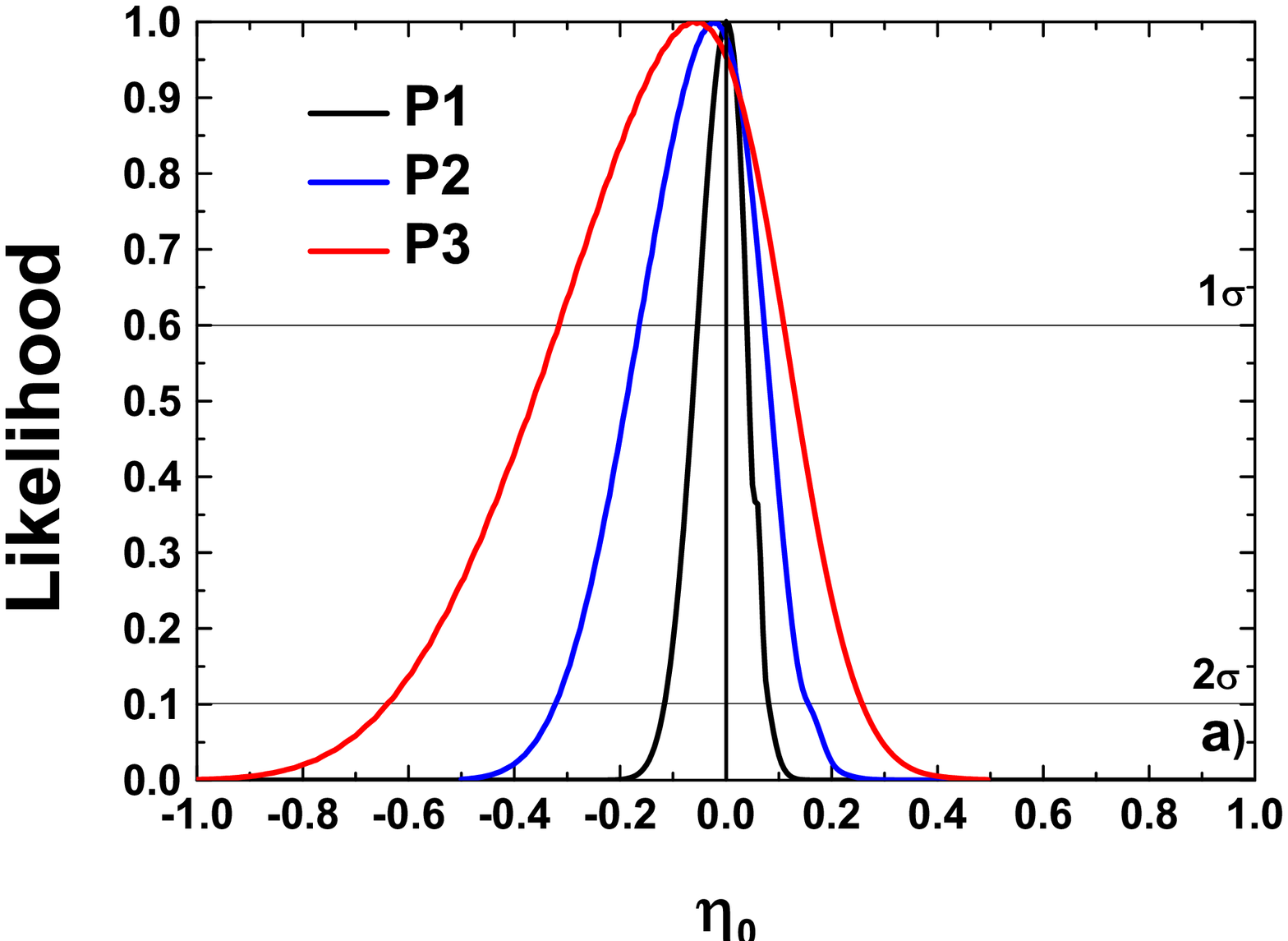}
\hspace{0.3cm}
\includegraphics[width=0.3\textwidth]{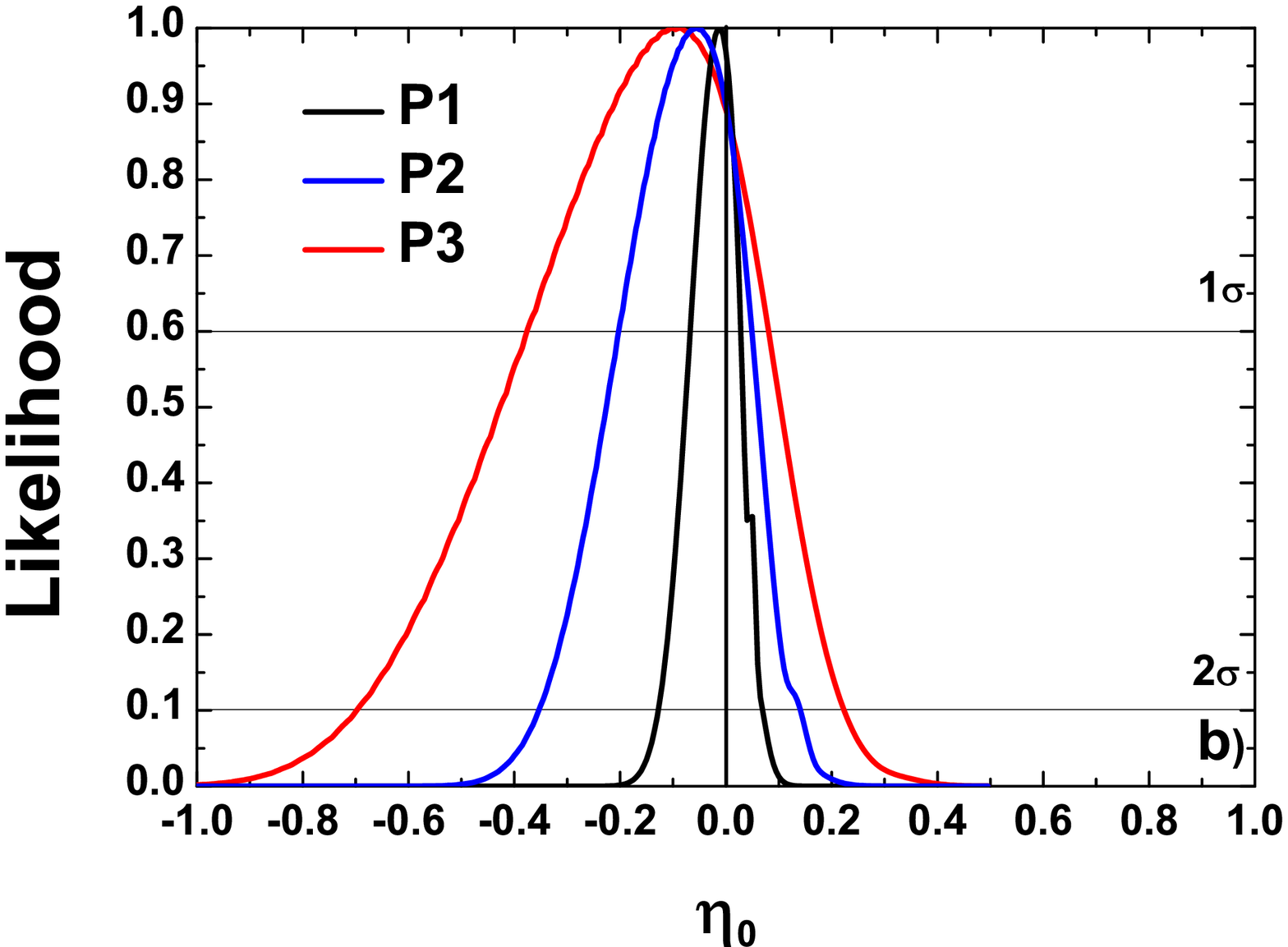}
\hspace{0.3cm}
\includegraphics[width=0.3\textwidth]{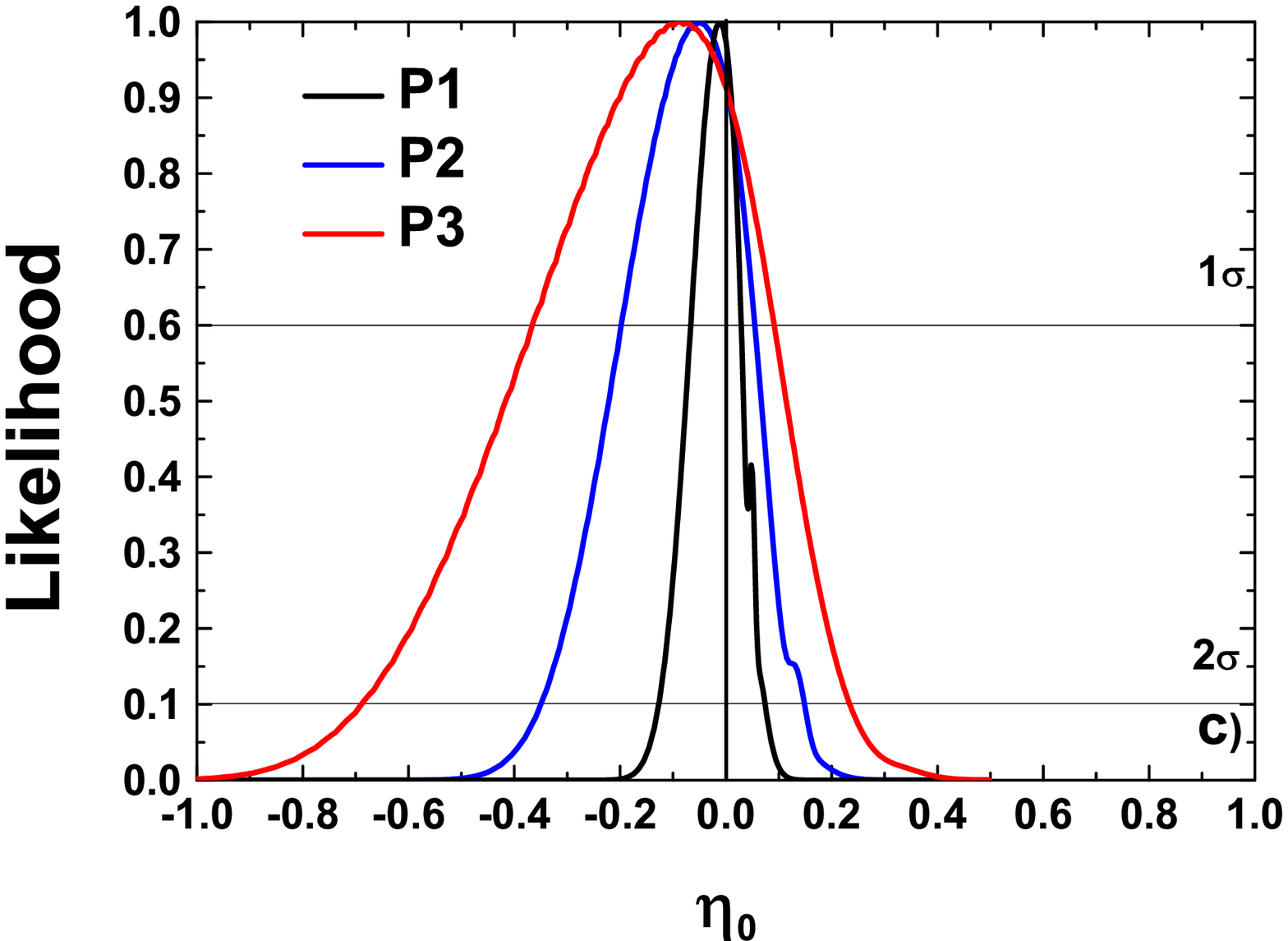}
\caption{ The likelihood functions from our analyses. In Figs.(a), (b) and (c) we plot the results by using 600, 900 and 1200 GWs simulated. In all figures the black, red and blue curves are for P1, P2 and P3 $\eta(z)$ functions.
}
\end{figure*}

\section{Luminosity distance from Gravitation Wave sources}
In this section, we briefly introduce the concept of standard siren and summarize the method
of~\cite{Cai:2016sby} employed to simulate observations of GW standard sirens from the ET. The data produced in~\cite{Cai:2016sby} has been used to forecast the constraints of cosmological
parameters such as $H_0$, $\Omega_m$ and the equation of state $w$ (see also~\cite{Cai:2017plb}).

The Einstein Telescope is the third generation of the ground based GW detector. As proposed by the design document, it consists of three collocated underground detectors, each with 10 km arm and with a 60$^\circ$ opening angle. The ET is envisaged to be ten times more sensitive in amplitude
than the advanced ground-based detectors, covering the frequency range of $1-10^4$ Hz. Here we
use ET to simulate the GW observations. As for the sources' redshifts, the binary merger of a NS with either a NS or BH is considered to be the progenitor of a short and intense burst of $\gamma$-rays (SGRB), an EM counterpart like SGRB can provide the redshift information if the host galaxy of the event can be pinpointed. The expected rates of BNS and BHNS detections per year for the ET~\footnote{\url{http://www.et-gw.eu/}} are about the order $10^3-10^7$. However, only a small fraction ($10^{-3}$) is expected to have
the observation of SGRB. In this work we take the detection rate in the middle range ($10^5$), thus the order $10^2$ events with SGRB per year.

For the waveform of GW, we also apply the stationary phase approximation to compute the Fourier transform $\mathcal{H}(f)$ of the time domain waveform $h(t)$,
\begin{align}
\mathcal{H}(f)=\mathcal{A}f^{-7/6}\exp[i(2\pi ft_0-\pi/4+2\psi(f/2)-\varphi_{(2.0)})],
\label{equa:hf}
\end{align}
where the Fourier amplitude $\mathcal{A}$ is given by
\begin{align}
\mathcal{A}=&~~\frac{1}{d_L}\sqrt{F_+^2(1+\cos^2(\iota))^2+4F_\times^2\cos^2(\iota)}\nonumber\\
            &~~\times \sqrt{5\pi/96}\pi^{-7/6}\mathcal{M}_c^{5/6}.
\label{equa:A}
\end{align}
Here $\mathcal{M}_c$ denotes the observed chirp mass, $d_L$ is the luminosity distance which plays the most important role in our purpose. The definition of the beam pattern function $F$ for ET and the phase parameters such as the angle of inclination $\iota$, the functions $\psi$ and $\varphi_{(2.0)}$ can be found in~\cite{Cai:2016sby,Zhao:2010sz}.

Since it is expected that SGRBs are strongly beamed~\cite{Abdo:2009zza,Nakar:2005bs,Rezzolla:2011da}, a coincident observation of the SGRB implies that the binary was orientated nearly face on, i.e., $\iota\approx0$. In fact the maximal inclination is about $\iota=20^\circ$; however, averaging the Fisher matrix over the inclination $\iota$ and the polarization $\psi$ with the constraint $\iota<20^\circ$ is approximately the same as taking $\iota=0$~\cite{Li:2013lza}. Therefore, when we simulate the GW source we can take $\iota=0$ and the Fourier amplitude $\mathcal{A}$ in Eq.~(\ref{equa:A}) will not then depend on the polarization angle $\psi$.

Using the Fisher matrix, we can calculate the uncertainty of the measurement of luminosity distance from the detector. We simulate  600, 900 and 1200 GW events with the redshift information provided by the SGRB (see red open circles in Figs. 1a, 1b and 1c). The fiducial model we choose is the concordant model based on the most recent {\it Plank} results~\cite{Ade:2015xua}:
\begin{align}
h_0=0.678,~~\Omega_m=0.308,~~\Omega_K=0,~~w=-1,
\label{ini}
\end{align}
here $H_0=100h_0$km\,s$^{-1}$Mpc$^{-1}$.
The $d_L-z$ data sets from ET can extend the reshifts range up to $z\sim 5$, which can easily cover the redshift range of the SGL system. Thus it is  suitable to combine these two data sets to test the CDDR.

We use the Gaussian process method to obtain luminosity distances and their errors for each SGL system from the mock GW data (see black filled circles in figs. 1a, 1b and 1c). The luminosity distance errors from 1200 and 900 simulated points are around $15\%$ and $30\%$, respectively, smaller than those from 600 points. The Gaussian process~\footnote{\url{http://www.gaussianprocess.org/gpml/}} is designed to use the supervised learning to build a model (or function) from the training data and then use the model to forecast the new samples.  The distribution over functions  provided by GP is suitable to describe the observed data. At each point $z$, the reconstructed function $f(z)$ is also a Gaussian distribution with a mean value and Gaussian error. The functions at different points $z$ and $\tilde{z}$ are related by a covariance function $k(z,\tilde{z})$, which only depends on a set of hyperparameters $\ell$ and $\sigma_f$. Here $\ell$ gives a measure of the coherence length of the correlation in $x$-direction and $\sigma_f$ denotes the overall amplitude of the correlation in the $y$-direction. Both of them will be optimized by GP with the observed data set.
In contrast to actual parameters, GP does not specify the form of the reconstructed function. Instead it characterizes the typical changes of the function. According to~\cite{Seikel:2013fda}, we choose the kernel function as the mat\'{e}rn ($\nu=9/2$) form
\begin{align}
k(z,\tilde z) = &{\sigma _f}^2\exp\left( - \frac{{3\left| {z - \tilde z} \right|}}{\ell }\right) \times\Big[1 + \frac{{3\left| {z - \tilde z} \right|}}{\ell } + \nonumber \\
&\frac{{27{{(z - \tilde z)}^2}}}{{7{\ell ^2}}}+ \frac{{18{{\left| {z - \tilde z} \right|}^3}}}{{7{\ell ^3}}} + \frac{{27{{(z - \tilde z)}^4}}}{{35{\ell ^4}}}\Big]\,.
\end{align}
The GP method has been used in many works~\cite{Holsclaw:2010nb,Holsclaw:2010sk,Holsclaw:2011wi,Gonzalez:2016lur}, We use the GaPP code developed in~\cite{Seikel:2012uu} to derive our results. For detailed description of the Gaussian process, one can refer to~\cite{Seikel:2012uu}, see also \cite{Cai:2015zoa, Cai:2015pia,Cai:2016vmn}. The reconstructed $d_L-z$ points from simulated GW data for each SGL system are plotted in Fig.~\ref{fig:plotdl}.

%\begin{table*}[ht]
%\caption{ The ratio between the 1$\sigma$ error widths of the results on $\eta_0$ from Refs.\cite{15,16} and the ours. The quantities between parenthesis correspond to results by using luminosity distance of GW sources obtained via Gaussian Process from 600, 900 and 1200 simulated points.}
%\label{likelihood1}
%\begin{center}
%\par
%\begin{tabular}{|c|c|c|c|}
%\hline
%\ \ Reference  & \  P1   & P2 &   P3  \\
%\hline
%Method A \cite{15}  & (1.50, 1.50, 1.50)  & (0.50 , 0.50, 0.50) & (1.38, 1.38, 1.38)  \\
%\hline
%Method B \cite{15} & (0.60, 0.60, 0.60) & (0.32, 0.32, 0.32) &  (1.01, 1.01, 1.01) \\
%\aline
%\cite{16} & (1.17, 1.17, 1.17) & (1.81, 1.81, 1.81) &  (1.92, 1.92, 1.92) \\
%\hline
%\end{tabular}
%\end{center}
%\end{table*}

\section{Data}

The   angular diameter distance data  used in our analyses were reported in Ref.~\cite{Cao} for 118 SGL systems  observed by the Sloan Lens ACS survey (SLACS), BOSS Emission-Line Lens Survey (BELLS), Lenses Structure and Dynamics Survey  (LSD) and  Strong Legacy Survey SL2S surveys. In this compilation, the 118 lens and sources are in the redshift ranges: $0.075 \leq z_l \leq 1.004$ and  $0.20 \leq z_s \leq 3.60$. Since several studies have shown that slopes of density profiles of individual galaxies show a non-negligible scatter from the singular isothermal sphere \cite{SIS}, we consider the approach of \cite{Cao} to describe the lensing systems called Power Law model. In the power law model (PLaw),  a spherically symmetric mass distribution in lensing galaxies is  assumed to obey a more general power-law index $\gamma $, $\rho \propto r^{-\gamma}$. By using the PLaw model, the Einstein radius is:
\begin{equation}
 \label{Einstein}
\theta_E =   4 \pi
\frac{\sigma_{ap}^2}{c^2} \frac{D_{ls}}{D_s} \left(
\frac{\theta_E}{\theta_{ap}} \right)^{2-\gamma} f(\gamma),
\end{equation}
where $\sigma_{ap}$ is the  stellar velocity dispersion inside the aperture of size $\theta_{ap}$ and
\begin{eqnarray} \label{f}
f(\gamma) &=& - \frac{1}{\sqrt{\pi}} \frac{(5-2 \gamma)(1-\gamma)}{3-\gamma} \frac{\Gamma(\gamma - 1)}{\Gamma(\gamma - 3/2)}\nonumber\\
          &\times & \left[ \frac{\Gamma(\gamma/2 - 1/2)}{\Gamma(\gamma / 2)} \right]^2.
\end{eqnarray}
Therefore:
\begin{equation}
\label{NewObservable}
 D=D_{A_{ls}}/D_{A_{s}} = \frac{c^2 \theta_E }{4 \pi \sigma_{ap}^2} \left( \frac{\theta_{ap}}{\theta_E} \right)^{2-\gamma} f^{-1}(\gamma).
\end{equation}
As one may see, the distribution becomes a SIS for $ \gamma = 2$. All  relevant information necessary to obtain $D$ in Eq.~(\ref{NewObservable}) for both models and perform our fit are in Table 1 of \cite{Cao}. In our analyses (see next section), we marginalize over $\gamma$ by using the following flat prior: $\gamma$: $1.5 < \gamma < 3.0$. Also following \cite{Cao}, we replace $\sigma_{ap}$ by $\sigma_0$ in the PLAW model.

In order to test the ability of the gravitational wave observations to impose limits  on possible departures from the CDDR, we consider the following CDDR parameterisations:

\begin{enumerate}
\item P1: $\eta(z)=1+\eta_0 z$;

\item P2: $\eta(z)=1+\eta_0 z/(1+z)$;

\item P3: $\eta(z)=1+\eta_0\ln(1+z)$.

\end{enumerate}
%This $\eta(z)$ functions were used recently in test involving SGL, SNe Ia and GRB, what able us to compare directly the results.

\section{Analyses and results}

The constraints on the $\eta_0$ parameter are derived by evaluating the likelihood distribution
function, ${\cal{L}} \propto e^{-\chi^{2}/2}$, with
\begin{eqnarray}
\chi^{2} & = & \sum_{i}^{118}\frac{\left[\frac{(1+z_{s_i})\eta(z_{s_i})}{(1+z_{l_i})\eta(z_{l_i})}- (1-D_i)\frac{D_{L_{s_i}}}{D_{L_{l_i}}}\right]^2}{\sigma_i^2}
\end{eqnarray}
where $\sigma_i^2$ stands for the statistical errors associated to the $D_L(z)$ of the GWs sources and  the gravitational lensing  observations and it is obtained by using standard propagation errors techniques. For the gravitational lensing error one may show that:
\begin{equation} \label{uncertainty}
\sigma_D = D \sqrt{4 (\delta \sigma_{0})^2 + (1-\gamma)^2 (\delta \theta_E)^2}\;,
\end{equation}

Our results for $\eta_0$ by using the 600, 900 and 1200 simulated luminosity distance data are plotted in figures (2a), (2b) and (2c). In all figures, the  black, red and blue solid lines  correspond to P1, P2 and P3, respectively. The black solid horizontal lines correspond to 1 and 2$\sigma$ (C.L.).  As one may see, the analyses are  consistent with the CDRR validity within 1$\sigma$ (C.L.).

At this point, it is interesting to compare our forecast results with some actual tests involving the angular diameter distances from SGL systems. In Ref.~\cite{16} the authors combined the 118 SGL measurements with luminosity distances from the Union2.1 compilation and the latest Gamma-Ray Bursts (GRBs) data. Two cosmological model-independent methods were considered to associate the redshifts of SNe Ia and GRBs with the redshits of the lens and source of observed from SGL systems: In method {\bf A} the SNe Ia + GRBs data with $|z_{SNe/GRB}-z_l| < 0.005$ and $|z_{SNe/GRB}-z_s| < 0.005$ were binned; in method {\bf B} a reconstruction of the luminosity distance  with the smoothing method by combining it with the Cross Statistic was performed. Other analysis were also performed in Ref.~\cite{15}, by fitting the luminosity distance of SNe Ia and GRBs with a second degree polynomial function.

By comparing the results at 1$\sigma$ c.l., we obtain   error bars 40\% smaller  than that from the method {\bf A} of the Ref.\cite{16}, regardless the $\eta(z)$ functions considered. The $1\sigma$ confidence level of each $\eta(z)$ function is 0.035,  By considering  the method {\bf B} of this same reference,  we obtain  larger error bars  when the P1 and P2 functions are considered, more precisely, 40\% and  30\%, respectively. For the P3 function the error bars are similar. Finally, by considering our results and those from Ref.\cite{15}, we obtain that our error bars are 20\%, 80\% and 90\% smaller when the P1, P2 and P3 functions are considered, respectively. The  results did not depend on the number GW simulated,  showing that the errors of the SGL systems dominate the fits. As one may see, in most case tighter limits are obtained by using luminosity distance of GW sources.  As commented earlier, the fundamental advantage here is not only using different set data for the luminosity distances at high redshifts but also the insensitive of the GWs to non-conservation of the number of photons.

\section{Conclusions}

The direct observations of gravitational waves made by LIGO/Virgo observatories opened up a new window to observational cosmology. More precisely, for this kind of event it is possible to measure the luminosity distance from waveform and amplitude of the gravitational waves observations, being this distance insensitive to a possible non-conservation of the number of photons, unlike other sources such as SNe Ia and GRBs.

In this work, we have used simulations based on the Einstein telescope and tested the potentialities of future measurements of the luminosity distances from gravitational waves sources. We have place limits on possible departures from the cosmic distance duality relation (CDDR) at high redshifts ($z\approx 3.6$) jointly with current estimates of the angular diameter distances from strong gravitational lensing systems. Simulating 600, 900 and 1200 events, we obtained the luminosity distances for each SGL system via Gaussian processes and put limits on $\eta(z)$ functions by considering three different parameterisations. %(see sections II and IV): $\eta(z)=1+\eta_0 z$, $\eta(z)=1+\eta_0 z/(1+z)$ and $\eta(z)=+\eta_0\ln(1+z)$ (if $\eta_0=0$ the validity is verified).

By comparing our results with previous ones (e.g., \cite{15,16}) which reported tests at high redshifts using current data of SNe Ia, GRBs and SGL systems, we have obtained that future measurements of the luminosity distances of gravitational waves sources will be at least competitive with the current analyses. However, unlike SNe  Ia and GRBs observations, distance measurements from GWs observations are insensitive  to non-conservation of the number of photons  and, then, a non-standard physical can be identified if possible departure from the CDDR validity is observationally verified.

\section*{Acknowledgments}
TY and BH are supported by the Beijing Normal University
Grant under the reference No. 312232102 and by the National Natural Science Foundation of China Grants No. 210100088 and No. 210100086. TY
is also supported by China Postdoctoral Science Foundation under Grants No. 2017M620662. BH is also partially supported
by the Chinese National Youth Thousand Talents Program under the reference No. 110532102 and the Fundamental Research
Funds for the Central Universities under the reference No.310421107.
RFLH acknowledges financial support from INCT-A and CNPq (No. 478524/2013-7, 303734/2014-0). R. F. L. Holanda thanks to Jailson Alcaniz by valuable suggestions.


\begin{thebibliography}{99}% Produces the bibliography via BibTeX.

\bibitem{Hinshaw:2012aka}
  G.~Hinshaw {\it et al.} [WMAP Collaboration],
 Nine-Year Wilkinson Microwave Anisotropy Probe (WMAP) Observations: Cosmological Parameter Results,
  Astrophys.\ J.\ Suppl.\  {\bf 208}, 19 (2013)
%  doi:10.1088/0067-0049/208/2/19
  [arXiv:1212.5226 [astro-ph.CO]].

\bibitem{Ade2015}
  P.~A.~R.~Ade {\it et al.} [Planck Collaboration],
Planck 2015 results. XIII. Cosmological parameters,
  Astron.\ Astrophys.\  {\bf 594}, A13 (2016)
  doi:10.1051/0004-6361/201525830
  [arXiv:1502.01589 [astro-ph.CO]].


\bibitem{Suzuki2012} N. Suzuki N. et al., The Hubble Space Telescope Cluster Supernova Survey. V. Improving the Dark-energy Constraints above $z > 1$ and Building an Early-type-hosted Supernova Sample,  ApJ, {\bf 85}, 746 (2012).

\bibitem{Betoule2014}
  M.~Betoule {\it et al.} [SDSS Collaboration],
  Improved cosmological constraints from a joint analysis of the SDSS-II and SNLS supernova samples, Astron.\ Astrophys.\  {\bf 568}, A22 (2014)
    [arXiv:1401.4064 [astro-ph.CO]].


\bibitem{Cole05}
S.~Cole {\it et al.} [2dFGRS Collaboration],The 2dF Galaxy Redshift Survey: Power-spectrum analysis of the final dataset and cosmological implications, Mon.\ Not.\ Roy.\ Astron.\ Soc., {\bf 362},505, (2005)

\bibitem{Eisenstein:2005}
  D.~J.~Eisenstein {\it et al.} [SDSS Collaboration],
  Detection of the Baryon Acoustic Peak in the Large-Scale Correlation Function of SDSS Luminous Red Galaxies,
  Astrophys.\ J.\  {\bf 633}, 560 (2005)
  [astro-ph/0501171].
\bibitem{Percival10} Percival W. J., Reid B. A., Eisenstein D. J., {\it et al.}, Baryon Acoustic Oscillations in the Sloan Digital Sky Survey Data Release 7 Galaxy Sample,  Mon.\ Not.\ Roy.\ Astron.\ Soc., {\bf 401}, 2148 (2010).

\bibitem{Carvalho:2017tuu}
  G.~C.~Carvalho, A.~Bernui, M.~Benetti, J.~C.~Carvalho and J.~S.~Alcaniz,Measuring the transverse baryonic acoustic scale from the SDSS DR11 galaxies,
  arXiv:1709.00271 [astro-ph.CO].


\bibitem{Weinberg2013}D. H. Weinberg et al., Observational probes of cosmic acceleration,  Phys. Rep., {\bf 530}, 87 (2013).

\bibitem{Maartens:2011yx}
  R.~Maartens,
  Is the Universe homogeneous?,
  Phil.\ Trans.\ Roy.\ Soc.\ Lond.\ A {\bf 369}, 5115 (2011)
 % doi:10.1098/rsta.2011.0289
  [arXiv:1104.1300 [astro-ph.CO]].

\bibitem{Clarkson2012}Clarkson C., Establishing homogeneity of the universe in the shadow of dark energy, Comptes rendus - Physique, {\bf 13}, 682 (2012).



\bibitem{Bengaly:2015xkw}
  C.~A.~P.~Bengaly, A.~Bernui, J.~S.~Alcaniz and I.~S.~Ferreira, Probing cosmological isotropy with Planck Sunyaev-Zeldovich galaxy clusters,
  Mon.\ Not.\ Roy.\ Astron.\ Soc.\  {\bf 466}, no. 3, 2799 (2017)
%  doi:10.1093/mnras/stw3233
  [arXiv:1511.09414 [astro-ph.CO]].

\bibitem{Bengaly:2016wto}  C.~A.~P.~Bengaly, A.~Bernui and J.~S.~Alcaniz, Constraining cosmic isotropy with type Ia supernovae,
  arXiv:1602.01389 [astro-ph.CO].

\bibitem{Goncalves:2017dzs}
  R.~S.~Goncalves, G.~C.~Carvalho, C.~A.~P.~Bengaly, J.~C.~Carvalho, A.~Bernui, J.~S.~Alcaniz and R.~Maartens,
 Cosmic homogeneity: a spectroscopic and model-independent measurement,
  arXiv:1710.02496 [astro-ph.CO].

\bibitem{Webb1999}
  J.~K.~Webb, V.~V.~Flambaum, C.~W.~Churchill, M.~J.~Drinkwater and J.~D.~Barrow,
  Evidence for time variation of the fine structure constant,  Phys.\ Rev.\ Lett.\  {\bf 82}, 884 (1999)
  [astro-ph/9803165].
  %%CITATION = doi:10.1103/PhysRevLett.82.884;%%
  %519 citations counted in INSPIRE as of 25 Oct 2017

\bibitem{Webb2013}Webb M. T. et al., Does the fine structure constant vary? A third quasar absorption sample consistent with varying $\alpha$, Astrophys. Space Science, {\bf 283}, 577 (2003)

\bibitem{Holanda2016a} R.~F.~L.~Holanda, S.~J.~Landau, J.~S.~Alcaniz, I.~E.~Sanchez G. and V.~C.~Busti,
 Constraints on a possible variation of the fine structure constant from galaxy cluster data,
  JCAP {\bf 1605}, no. 05, 047 (2016)
   [arXiv:1510.07240 [astro-ph.CO]].

\bibitem{Holanda2016b} R.~F.~L.~Holanda, V.~C.~Busti, L.~R.~Colaco, J.~S.~Alcaniz and S.~J.~Landau,
  Galaxy clusters, type Ia supernovae and the fine structure constant,
  JCAP {\bf 1608}, no. 08, 055 (2016)
   [arXiv:1605.02578 [astro-ph.CO]].
\bibitem{Luzzi2009} Luzzi G. {\it et al.}, Redshift Dependence of the CMB Temperature from S-Z Measurements, Astrophys. J., {\bf 705}, 1122 (2009).
\bibitem{Luzzi2015}G.~Luzzi, R.~T.~Genova-Santos, C.~J.~A.~P.~Martins, M.~De Petris and L.~Lamagna,
 Constraining the evolution of the CMB temperature with SZ measurements from Planck data,  JCAP {\bf 1509}, no. 09, 011 (2015)
  [arXiv:1502.07858 [astro-ph.CO]].
  %%CITATION = doi:10.1088/1475-7516/2015/09/011;%%
  %17 citations counted in INSPIRE as of 25 Oct 2017
\bibitem{Hurier2014} G.~Hurier, N.~Aghanim, M.~Douspis and E.~Pointecouteau,
  Measurement of the $T_{\rm CMB}$ evolution from the Sunyaev-Zel'dovich effect,  Astron.\ Astrophys.\  {\bf 561}, A143 (2014)
  [arXiv:1311.4694 [astro-ph.CO]].
\bibitem{Etherington1933}Etherington I. M. H., On the definition of distance in general relativity,  Phil. Mag, {\bf 15}, 761 (1933).
\bibitem{Ellis2007} Ellis G. F. R., Relativistic cosmology, Gen. Rel. Grav., {\bf 39}, 1047 (2007).
\bibitem{Uzan2004}
  J.~P.~Uzan, N.~Aghanim and Y.~Mellier,
 The Distance duality relation from x-ray and SZ observations of clusters,
  Phys.\ Rev.\ D {\bf 70}, 083533 (2004).
  [astro-ph/0405620].
\bibitem{Santana:2017zvy}
  L.~T.~Santana, M.~O.~Calvao, R.~R.~R.~Reis and B.~B.~Siffert, How does light move in a generic metric-affine background?,
  Phys.\ Rev.\ D {\bf 95}, no. 6, 061501 (2017).
%  doi:10.1103/PhysRevD.95.061501
  [arXiv:1703.10871 [gr-qc]].

\bibitem{basset2004}  B.~A.~Bassett and M.~Kunz,
 Cosmic distance-duality as a probe of exotic physics and acceleration,
  Phys.\ Rev.\ D {\bf 69}, 101305 (2004)
  [astro-ph/0312443].
	
\bibitem{avg2009}
  A.~Avgoustidis, L.~Verde and R.~Jimenez, Consistency among distance measurements: transparency, BAO scale and accelerated expansion,
  JCAP {\bf 0906}, 012 (2009)
  [arXiv:0902.2006 [astro-ph.CO]].
	
\bibitem{avg2010}
  A.~Avgoustidis, C.~Burrage, J.~Redondo, L.~Verde and R.~Jimenez,
 Constraints on cosmic opacity and beyond the standard model physics from cosmological distance measurements,
  JCAP {\bf 1010}, 024 (2010)
  [arXiv:1004.2053 [astro-ph.CO]].
\bibitem{jac2010}
  J.~Jaeckel and A.~Ringwald,
 The Low-Energy Frontier of Particle Physics,
  Ann.\ Rev.\ Nucl.\ Part.\ Sci.\  {\bf 60}, 405 (2010)
  [arXiv:1002.0329 [hep-ph]].
\bibitem{Hees2014} A.~Hees, O.~Minazzoli and J.~Larena, Breaking of the equivalence principle in the electromagnetic sector and its cosmological signatures,  Phys.\ Rev.\ D {\bf 90}, 124064 (2014) [arXiv:1406.6187 [astro-ph.CO]].
\bibitem{HolandaBarros2016}
  R.~F.~L.~Holanda and K.~N.~N.~O.~Barros,
 Searching for cosmological signatures of the Einstein equivalence principle breaking,
  Phys.\ Rev.\ D {\bf 94}, no. 2, 023524 (2016) [arXiv:1606.07923 [astro-ph.CO]].
\bibitem{HolandaSaulo2017} R.~F.~L.~Holanda and S.~H.~Pereira,
  Can galaxy clusters, type Ia supernovae and cosmic microwave background rule out a class of modified gravity theories?,
  Phys.\ Rev.\ D {\bf 94}, no. 10, 104037 (2016)
  [arXiv:1610.01512 [astro-ph.CO]].
\bibitem{Holandasimoni2017} R.~F.~L.~Holanda, S.~H.~Pereira and S.~Santos da Costa,
  Searching for deviations from the General Relativity Theory with gas mass fraction of galaxy clusters and complementary probes,
  Phys.\ Rev.\ D {\bf 95}, no. 8, 084006 (2017)
  [arXiv:1612.09365 [astro-ph.CO]].

\bibitem{Belgacem:2017ihm}
  E.~Belgacem, Y.~Dirian, S.~Foffa and M.~Maggiore,
  %``The gravitational-wave luminosity distance in modified gravity theories,''
  arXiv:1712.08108 [astro-ph.CO].
  %%CITATION = ARXIV:1712.08108;%%
  %7 citations counted in INSPIRE as of 17 Apr 2018

\bibitem{HGA2012}
  R.~F.~L.~Holanda, R.~S.~Goncalves and J.~S.~Alcaniz,
  A test for cosmic distance duality,
  JCAP {\bf 1206}, 022 (2012)
  [arXiv:1201.2378 [astro-ph.CO]].


\bibitem{5}
  R.~F.~L.~Holanda, J.~A.~S.~Lima and M.~B.~Ribeiro,
  Testing the Distance-Duality Relation with Galaxy Clusters and Type Ia Supernovae,
  Astrophys.\ J.\  {\bf 722}, L233 (2010)
  [arXiv:1005.4458 [astro-ph.CO]].
\bibitem{6}
  S.~Santos-da-Costa, V.~C.~Busti and R.~F.~L.~Holanda,
Two new tests to the distance duality relation with galaxy clusters,
  JCAP {\bf 1510}, no. 10, 061 (2015)
  [arXiv:1506.00145 [astro-ph.CO]].
\bibitem{7}R.~S.~Goncalves, J.~S.~Alcaniz, J.~C.~Carvalho and R.~F.~L.~Holanda,
  Forecasting constraints on the cosmic duality relation with galaxy clusters,
  Phys.\ Rev.\ D {\bf 91}, no. 2, 027302 (2015)
  [arXiv:1306.6644 [astro-ph.CO]].
\bibitem{8}Meng X.-L., Zhang T.-J.,  Zhan H., Morphology of Galaxy Clusters: A Cosmological Model-Independent Test of the Cosmic Distance-Duality Relation, ApJ, {\bf 745}, 98 (2012).
\bibitem{9}R.~Nair, S.~Jhingan and D.~Jain, Observational Cosmology And The Cosmic Distance Duality Relation, JCAP {\bf 1105}, 023 (2011)
[arXiv:1102.1065 [astro-ph.CO]].
\bibitem{10}Wu P., Li Z., Liu X., Yu H., Cosmic distance-duality relation test using type Ia supernovae and the baryon acoustic oscillation , Phys. Rev. D, {\bf 92}, 023520 (2011).
\bibitem{11} X.~Yang, H.~R.~Yu and T.~J.~Zhang, Constraining smoothness parameter and the DD relation of Dyer-Roeder equation with supernovae,
  JCAP {\bf 1306}, 007 (2013)
  [arXiv:1305.6989 [astro-ph.CO]].
\bibitem{12}  S.~Jhingan, D.~Jain and R.~Nair, Observational cosmology and the cosmic distance-duality relation,
  J.\ Phys.\ Conf.\ Ser.\  {\bf 484}, 012035 (2014)
  [arXiv:1403.2070 [gr-qc]].
\bibitem{13}  A.~Rana, D.~Jain, S.~Mahajan, A.~Mukherjee and R.~F.~L.~Holanda,
  Probing the cosmic distance duality relation using time delay lenses,
  JCAP {\bf 1707}, no. 07, 010 (2017)
  [arXiv:1705.04549 [astro-ph.CO]].
\bibitem{13a}
  K.~Liao, Z.~Li, S.~Cao, M.~Biesiada, X.~Zheng and Z.~H.~Zhu,
 The Distance Duality Relation From Strong Gravitational Lensing,
  Astrophys.\ J.\  {\bf 822}, no. 2, 74 (2016)
  [arXiv:1511.01318 [astro-ph.CO]].

\bibitem{15}
  R.~F.~L.~Holanda, V.~C.~Busti, F.~S.~Lima and J.~S.~Alcaniz,
 Probing the distance-duality relation with high-$z$ data,
  JCAP {\bf 1709}, no. 09, 039 (2017)
  [arXiv:1611.09426 [astro-ph.CO]].
\bibitem{16}  X.~Fu and P.~Li,
Testing the distance-duality relation from strong gravitational lensing, type Ia supernovae and gamma-ray bursts data up to redshift $z\sim3.6$,
  Int.\ J.\ Mod.\ Phys.\ D {\bf 26}, no. 9, 1750097 (2017)
  [arXiv:1702.03626 [gr-qc]].
\bibitem{grb}  M.~Demianski, E.~Piedipalumbo, D.~Sawant and L.~Amati,
 Cosmology with gamma-ray bursts: I. The Hubble diagram through the calibrated $E_{\rm p,i}$ - $E_{\rm iso}$ correlation,''
  Astron.\ Astrophys.\  {\bf 598}, A112 (2017)
   [arXiv:1610.00854 [astro-ph.CO]].
\bibitem{Cao}
  S.~Cao, M.~Biesiada, R.~Gavazzi, A.~Piarkowska and Z.~H.~Zhu,
  Cosmology With Strong-lensing Systems,
  Astrophys.\ J.\  {\bf 806}, 185 (2015)
  [arXiv:1509.07649 [astro-ph.CO]].
\bibitem{SIS} Barnabe P. et al., 2011, Two-dimensional kinematics of SLACS lenses: III. Mass structure and dynamics of early-type lens galaxies beyond $z \approx 0.1$, Mont. Not. Royal Astron. Soc., {\bf 415}, 2215 (2011).

\bibitem{TheLIGOScientific:2017qsa}
  B.~P.~Abbott {\it et al.} [LIGO Scientific and Virgo Collaborations],
  %``GW170817: Observation of Gravitational Waves from a Binary Neutron Star Inspiral,''
  Phys.\ Rev.\ Lett.\  {\bf 119}, no. 16, 161101 (2017)
%  doi:10.1103/PhysRevLett.119.161101
  [arXiv:1710.05832 [gr-qc]].

  \bibitem{Monitor:2017mdv}
  B.~P.~Abbott {\it et al.} [LIGO Scientific and Virgo and Fermi-GBM and INTEGRAL Collaborations],
  Gravitational Waves and Gamma-Rays from a Binary Neutron Star Merger: GW170817 and GRB 170817A,
  Astrophys.\ J.\  {\bf 848}, no. 2, L13 (2017)
  %doi:10.3847/2041-8213/aa920c
  [arXiv:1710.05834 [astro-ph.HE]].

     \bibitem{Diaz:2017uch}
  M.~C.~Diaz {\it et al.} [TOROS Collaboration],
  ``Observations of the first electromagnetic counterpart to a gravitational wave source by the TOROS collaboration,
  %Submitted to: Astrophys.J.
  [arXiv:1710.05844 [astro-ph.HE]].

  \bibitem{Cowperthwaite:2017dyu}
  P.~S.~Cowperthwaite {\it et al.},
 The Electromagnetic Counterpart of the Binary Neutron Star Merger LIGO/Virgo GW170817. II. UV, Optical, and Near-infrared Light Curves and Comparison to Kilonova Models,
  Astrophys.\ J.\  {\bf 848}, no. 2, L17 (2017)
%  doi:10.3847/2041-8213/aa8fc7

\bibitem{DiValentino:2017clw}
  E.~Di Valentino and A.~Melchiorri,
  First cosmological constraints combining Planck with the recent gravitational-wave standard siren measurement of the Hubble constant,
  arXiv:1710.06370 [astro-ph.CO].


\bibitem{Cai:2017cbj}
  R.~G.~Cai, Z.~Cao, Z.~K.~Guo, S.~J.~Wang and T.~Yang,
  The Gravitational-Wave Physics,''
  arXiv:1703.00187 [gr-qc].
  %%CITATION = doi:10.1093/nsr/nwx029;%%
  %14 citations counted in INSPIRE as of 11 Sep 2017

\bibitem{Schutz:1986gp}
  B.~F.~Schutz,
  Determining the Hubble Constant from Gravitational Wave Observations,
  Nature {\bf 323}, 310 (1986).
  %%CITATION = doi:10.1038/323310a0;%%
  %364 citations counted in INSPIRE as of 11 Sep 2017

\bibitem{Cai:2016sby}
  R.~G.~Cai and T.~Yang,
  Estimating cosmological parameters by the simulated data of gravitational waves from the Einstein Telescope,
  Phys.\ Rev.\ D {\bf 95}, no. 4, 044024 (2017)
  [arXiv:1608.08008 [astro-ph.CO]].
  %%CITATION = doi:10.1103/PhysRevD.95.044024;%%
  %6 citations counted in INSPIRE as of 11 Sep 2017

\bibitem{Cai:2017plb}
  R.~G.~Cai and T.~Yang,
  Standard sirens and dark sector with Gaussian process,
  arXiv:1709.00837 [astro-ph.CO].
  %%CITATION = ARXIV:1709.00837;%%

\bibitem{Zhao:2010sz}
  W.~Zhao, C.~Van Den Broeck, D.~Baskaran and T.~G.~F.~Li,
  Determination of Dark Energy by the Einstein Telescope: Comparing with CMB, BAO and SNIa Observations,
  Phys.\ Rev.\ D {\bf 83}, 023005 (2011)
  [arXiv:1009.0206 [astro-ph.CO]].
  %%CITATION = doi:10.1103/PhysRevD.83.023005;%%
  %40 citations counted in INSPIRE as of 11 Sep 2017

\bibitem{Abdo:2009zza}
  A.~A.~Abdo {\it et al.} [Fermi-LAT and Fermi GBM Collaborations],
  Fermi Observations of High-Energy Gamma-Ray Emission from GRB 080916C,
  Science {\bf 323}, 1688 (2009).
  doi:10.1126/science.1169101
  %%CITATION = doi:10.1126/science.1169101;%%
  %379 citations counted in INSPIRE as of 23 Aug 2016

\bibitem{Nakar:2005bs}
  E.~Nakar, A.~Gal-Yam and D.~B.~Fox,
  The Local Rate and the Progenitor Lifetimes of Short-Hard Gamma-Ray Bursts: Synthesis and Predictions for LIGO,
  Astrophys.\ J.\  {\bf 650}, 281 (2006)
  doi:10.1086/505855
  [astro-ph/0511254].
  %%CITATION = doi:10.1086/505855;%%
  %110 citations counted in INSPIRE as of 23 Aug 2016

\bibitem{Rezzolla:2011da}
  L.~Rezzolla, B.~Giacomazzo, L.~Baiotti, J.~Granot, C.~Kouveliotou and M.~A.~Aloy,
  The missing link: Merging neutron stars naturally produce jet-like structures and can power short Gamma-Ray Bursts,
  Astrophys.\ J.\  {\bf 732}, L6 (2011)
  doi:10.1088/2041-8205/732/1/L6
  [arXiv:1101.4298 [astro-ph.HE]].
  %%CITATION = doi:10.1088/2041-8205/732/1/L6;%%
  %170 citations counted in INSPIRE as of 23 Aug 2016

\bibitem{Li:2013lza}
  Tjonnie~G.~F.~Li,
  Extracting Physics from Gravitational Waves,
  Springer Theses (2015)
  doi:10.1007/978-3-319-19273-4
  %%CITATION = INSPIRE-1266133;%%


\bibitem{Ade:2015xua}
  P.~A.~R.~Ade {\it et al.} [Planck Collaboration],
Planck 2015 results. XIII. Cosmological parameters,
  Astron.\ Astrophys.\  {\bf 594}, A13 (2016)
  [arXiv:1502.01589 [astro-ph.CO]].
  %%CITATION = doi:10.1051/0004-6361/201525830;%%
  %3982 citations counted in INSPIRE as of 11 Sep 2017
  
\bibitem{Seikel:2013fda} 
  M.~Seikel and C.~Clarkson,
  Optimising Gaussian processes for reconstructing dark energy dynamics from supernovae,
  arXiv:1311.6678 [astro-ph.CO].
  %%CITATION = ARXIV:1311.6678;%%
  %31 citations counted in INSPIRE as of 23 Nov 2018  

\bibitem{Holsclaw:2010nb}
  T.~Holsclaw, U.~Alam, B.~Sanso, H.~Lee, K.~Heitmann, S.~Habib and D.~Higdon,
 Nonparametric Reconstruction of the Dark Energy Equation of State,
  Phys.\ Rev.\ D {\bf 82}, 103502 (2010)
  [arXiv:1009.5443 [astro-ph.CO]].
  %%CITATION = doi:10.1103/PhysRevD.82.103502;%%
  %46 citations counted in INSPIRE as of 21 Aug 2017

\bibitem{Holsclaw:2010sk}
  T.~Holsclaw, U.~Alam, B.~Sanso, H.~Lee, K.~Heitmann, S.~Habib and D.~Higdon,
 Nonparametric Dark Energy Reconstruction from Supernova Data,
  Phys.\ Rev.\ Lett.\  {\bf 105}, 241302 (2010)
  [arXiv:1011.3079 [astro-ph.CO]].
  %%CITATION = doi:10.1103/PhysRevLett.105.241302;%%
  %67 citations counted in INSPIRE as of 21 Aug 2017

\bibitem{Holsclaw:2011wi}
  T.~Holsclaw, U.~Alam, B.~Sanso, H.~Lee, K.~Heitmann, S.~Habib and D.~Higdon,
 Nonparametric Reconstruction of the Dark Energy Equation of State from Diverse Data Sets,
  Phys.\ Rev.\ D {\bf 84}, 083501 (2011)
  [arXiv:1104.2041 [astro-ph.CO]].

  \bibitem{Gonzalez:2016lur}
  J.~E.~Gonzalez, J.~S.~Alcaniz and J.~C.~Carvalho,
  ``Non-parametric reconstruction of cosmological matter perturbations,''
  JCAP {\bf 1604}, no. 04, 016 (2016)
  [arXiv:1602.01015 [astro-ph.CO]].

\bibitem{Seikel:2012uu}
  M.~Seikel, C.~Clarkson and M.~Smith,
Reconstruction of dark energy and expansion dynamics using Gaussian processes,
  JCAP {\bf 1206}, 036 (2012)
  [arXiv:1204.2832 [astro-ph.CO]].
  %%CITATION = doi:10.1088/1475-7516/2012/06/036;%%
  %61 citations counted in INSPIRE as of 21 Aug 2017



\bibitem{Cai:2015zoa}
  T.~Yang, Z.~K.~Guo and R.~G.~Cai,
Reconstructing the interaction between dark energy and dark matter using Gaussian Processes,
  Phys.\ Rev.\ D {\bf 91}, no. 12, 123533 (2015)
  [arXiv:1505.04443 [astro-ph.CO]].
  %%CITATION = doi:10.1103/PhysRevD.91.123533;%%
  %23 citations counted
  %in INSPIRE as of 15 Aug 2017


\bibitem{Cai:2016vmn}
  R.~G.~Cai, Z.~K.~Guo and T.~Yang,
 Dodging the cosmic curvature to probe the constancy of the speed of light,
  JCAP {\bf 1608}, no. 08, 016 (2016)
  [arXiv:1601.05497 [astro-ph.CO]].
  %%CITATION = doi:10.1088/1475-7516/2016/08/016;%%
  %4 citations counted in INSPIRE as of 27 Aug 2017

\bibitem{Cai:2015pia}
  R.~G.~Cai, Z.~K.~Guo and T.~Yang,
 Null test of the cosmic curvature using $H(z)$ and supernovae data,
  Phys.\ Rev.\ D {\bf 93}, no. 4, 043517 (2016)
  [arXiv:1509.06283 [astro-ph.CO]].
%%%The bibtem added by Tao Yang until this postion



\end{thebibliography}
\end{document}